\documentclass[aps,prl,twocolumn,superscriptaddress,showkeys,amsmath,amssymb]{revtex4-1}		
\usepackage{graphicx}
\usepackage{dcolumn}
\usepackage{bm}
\usepackage{hyperref}
\usepackage{subfig}

\usepackage[font=small,labelfont=bf,format=plain,justification=centerlast,labelsep=period]{caption}
\usepackage{txfonts}
\usepackage[usenames, dvipsnames]{color}

\newcommand{\ket}[1]{\left\vert #1 \right>}

\usepackage{verbatim}
\usepackage[normalem]{ulem}

\begin{document}

\preprint{APS/123-QED}

\title{Dynamical Casimir effect in stochastic systems: photon-harvesting through noise}

\author{Ricardo Rom\'an-Ancheyta}
\email{ancheyta6@gmail.com}
\affiliation{Instituto de Ciencias F\'isicas,
Universidad Nacional Aut\'onoma de M\'exico, Apartado Postal 48-3, 62251 Cuernavaca, Morelos, M\'exico}
\author{Ir\'an Ramos-Prieto}
\affiliation{Instituto Nacional de Astrof\'isica, \'Optica y
Electr\'onica, Calle Luis Enrique Erro 1, Santa Mar\'ia Tonantzintla, Puebla CP 72840, M\'exico}
\author{Armando Perez-Leija}
\affiliation{Max-Born-Institut, Max-Born-Stra{\ss}e 2A, 12489 Berlin, Germany}
\affiliation{Humboldt-Universit\"{a}t zu Berlin, Institut f\"{u}r Physik, AG Theoretische optik Photonik, Newtonstra{\ss}e 15, 12489 Berlin, Germany}
\author{Kurt Busch}
\affiliation{Max-Born-Institut, Max-Born-Stra{\ss}e 2A, 12489 Berlin, Germany}
\affiliation{Humboldt-Universit\"{a}t zu Berlin, Institut f\"{u}r Physik, AG Theoretische optik Photonik, Newtonstra{\ss}e 15, 12489 Berlin, Germany}
\author{Roberto de J. Le\'on-Montiel}
\email{roberto.leon@nucleares.unam.mx}
\affiliation{Instituto de Ciencias Nucleares, Universidad Nacional Aut\'onoma de
M\'exico,\\ Apartado Postal 70-543, 04510 Cd. Mx., M\'exico}
%

\begin{abstract}
We theoretically investigate the dynamical Casimir effect in a single-mode cavity endowed with a driven off-resonant mirror. We explore the dynamics of photon generation as a function of the ratio between the cavity mode and the mirror's driving frequency. Interestingly, we find that this ratio defines a threshold---which we referred to as a metal-insulator phase transition---between an exponential growth and a low photon production. The low photon production is due to Bloch-like oscillations that produce a strong localization of the initial vacuum state, thus preventing higher generation of photons. To break localization of the vacuum state, and enhance the photon generation, we impose a dephasing mechanism, based on dynamic disorder, into the driving frequency of the mirror. Additionally, we explore the effects of finite temperature on the photon production. Concurrently, we propose a classical analogue of the dynamical Casimir effect in engineered photonic lattices, where the propagation of classical light emulates the photon generation from the quantum vacuum of a single-mode tunable cavity.

\end{abstract}




\maketitle


{\em Introduction}.-
One of the most fundamental results of quantum theory is that vacuum space is not really empty.
Along these lines, in 1948, Casimir \cite{casimir1948} predicted that two parallel mirrors placed in empty space would experience an attractive force as a result of the spatial mismatch between the vacuum modes contained in the cavity and those outside of the mirrors. Remarkably, if the mirrors are allowed to move
non-adiabatically, the vacuum mode mismatch may occur in time rather than space. In such a situation, the cavity field does not remain in the vacuum state but gives rise to the generation of photons out of vacuum fluctuations \cite{moore1970,RevModPhysFNori}.
This fascinating phenomenon, termed dynamical Casimir effect (DCE), can be understood as a parametric amplification of vacuum fluctuations \cite{yablonovitch1989,schwinger1992,DodonovReview}. Indeed, it has been shown that a cavity field can be parametrically excited when the cavity length is periodically modulated \cite{ricardo2015,roberto2015}. Particularly, in the case where the cavity field is initially in the vacuum state, one can show that its evolution leads to a squeezed vacuum state \cite{ricardo2017,giacobino1992}, which---unlike a pure vacuum state---contains real photons \cite{breitenbach1997}.

To date, several experimental schemes to observe the DCE have been proposed \cite{lambrecht1996,uhlmann2004,braggio2005,kim2006,liberato2007,johansson2010,wilson2010}, but only a few have succeeded \cite{wilson2011,lahteenmaki2013}. The main limitation is because a non-negligible photon production can only be attained when the mirror's speed becomes  comparable to the speed of light. Consequently, observations of DCEs represent a very challenging task.
Clearly, of importance will be to identify equivalent systems upon which non-adiabatic changes of boundary conditions can be mapped to other physical variables. For instance, in Refs. \cite{wilson2011,lahteenmaki2013}, the equivalent action of a fast moving mirror is mimicked by an inductance variation of a superconducting quantum interference device (SQUID) controlled by a fast oscillating magnetic flux. Unlike actual mirrors, the inductance of a SQUID can be driven at high frequencies ($>10$ GHZ), which enables an experimentally detectable photon production.

In this paper, we put forward an experimental setup, based on semi-infinite waveguide arrays, in which the propagation of classical light emulates the generation of photons from the quantum vacuum of a single-mode tunable cavity. Using such waveguide configurations, we are able to emulate and explore the dynamics of photon generation as a function of the ratio between the fundamental cavity mode and the driving frequency. Interestingly, through this optical analogue, we find a threshold at which the exponentially increasing photon production abruptly drops down. This effect occurs due to the emergence of Bloch-like oscillations that produce strong localization of the initial vacuum state and prevents the generation of photons. In order to break such localization, and enhance the photon generation, we propose a dephasing mechanism based on dynamic disorder or noise. Finally, we explore the effects of finite temperature on the photon production and provide a proposal for its implementation.

{\em Dynamical Casimir effect}.-
We start by considering the e\-ff\-ective quantum Hamiltonian describing
the dynamics of the electromagnetic field contained in an ideal one-dimensional
cavity with a movable mirror, whose position is described by the function $q(t)$ \cite{Law1,Soff1988}: 
%

\begin{align}\label{:ham:eff:fock}
H_{\textit{\rm eff}}(t)&=\small{\sum}_k{\{}\omega^{\mathstrut}_k(t)a_k^\dagger a^{\mathstrut}_k+
\mathrm{i}\chi^{\mathstrut}_k(t)\big(a^{\dagger 2}_k-a^2_k\big){\}}\nonumber\\
&+\small{\sum}_{k,j, k\neq j }\frac{\mathrm{i}}{2}\mu^{\mathstrut}_{kj}(t)\lbrace a^{\dagger}_ka^\dagger_j
+a^\dagger_ka^{\mathstrut}_j-a^{\mathstrut}_ja^{\mathstrut}_k-a^\dagger_ja^{\mathstrut}_k\rbrace,
\end{align}
where $a^{\mathstrut}_k$ and $a^{\dagger }_k$ are the bosonic operators for the $k$-th field mode satisfying the commutation relation $[a^{\mathstrut}_k,a^\dagger_j]$=$\delta_{k,j}$. $\omega^{\mathstrut}_k(t)$=${k\pi}/{q(t)}$ is the instantaneous cavity frequency, 
$\chi^{\mathstrut}_k(t)$=${\dot{\omega}_k(t)}[{4\omega_k(t)}]^{-1}$ is an squeezing coefficient that multiplies the terms that create photon pairs from the vacuum state, and
$\mu^{\mathstrut}_{kj}(t)=(-1)^{k+j}2k\sqrt{{k}{j}}\dot{q}(t)[(j^2-k^2)q(t)]^{-1}$ represents the intermode interaction. Here the dot stands for the time derivative, and we have set $\hbar$=$c$=$1$ along with the dielectric permittivity. Notice that in static conditions, i.e. $\dot{q}(t)=0$, $H_{\rm eff}(t)$ reduces to a set of uncoupled harmonic oscillators with fixed frequency.

\begin{figure}[t!]
\includegraphics[width=7cm,height=5.5cm]{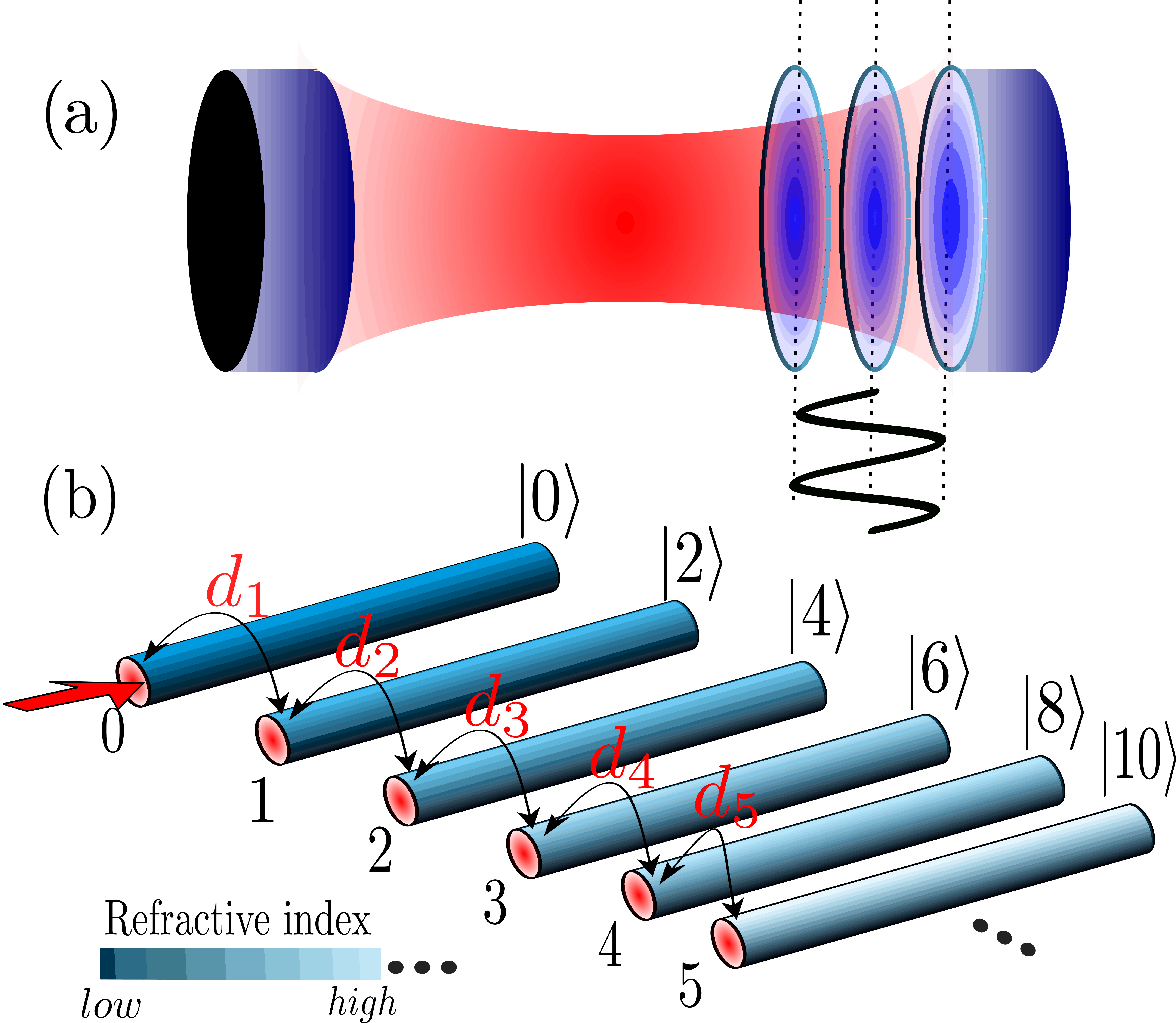}
\caption{
(a) Schematic representation of a non-stationary electromagnetic 
cavity in which the dynamical Casimir effect is manifested.
(b) Proposed semi-infinite squeezed-like waveguide array for the simulation of photon productions from vacuum.} 
\label{vibrating_cavity}
\vspace{-0.6cm}
\end{figure}

To the best of our knowledge there is no general analy\-tical solution to the corresponding Schr\"odinger equation for arbitrary
$q(t)$ in the above Hamiltonian. As a result, one must perform several approximations to derive a solution.
For instance, in Ref.~\cite{Klimov1}, the authors report approximated analytical solutions describing photon generation in cavities endowed with a movable mirror obeying sinusoidal trajectories and a resonance condition, where the mirror frequency is twice
the frequency of some unperturbed cavity mode. In such a scenario, the photon generation rate in the fundamental cavity mode rapidly reaches a constant value while the total number of created photons in all modes $\mathcal{N}_{tot}=\sum_{n=1}^\infty\mathcal{N}_n$ increases quadratically with time~\cite{DodonovReview}, where $\mathcal{N}_n$ is the average photon number in the $n$-th mode.
%

In what follows we consider a single-mode cavity [see Fig.~\ref{vibrating_cavity} (a) for illustration] 
such that the intermode interaction term in Eq.~\eqref{:ham:eff:fock} vanishes~\cite{RevModPhysFNori}.
In this single-mode regime the system is governed by an effective Hamiltonian that des\-cribes the DCE in absence of dissipation, namely $H_{\rm eff}(t)\approx\omega(t)a^\dagger a+{\rm i}\chi(t)(a^{\dagger 2}-a^2)$.
Furthermore, we assume a harmonic time-dependent frequency
$\omega(t)$=$\omega_0[1+\epsilon\sin(\nu t)]$, where $\omega_0$ is the
fundamental mode frequency when the mirrors are fixed, and $\epsilon$, $\nu$
are the amplitude and frequency of modulation, res\-pec\-tively.
Typically, the amplitude satisfies the condition $\epsilon\ll 1$, so that the squeezing parameter can be approximated as $\chi (t)\simeq (\epsilon\nu/4)\cos(\nu t)$ and the cavity frequency becomes fixed, that is, $\omega(t)\simeq\omega_0$~\cite{DodonovOneAtom}.

To explore the dynamics of the system in a general, off-resonance, regime we take $\nu$=$2\omega_{0}$+$K$, with $K$ representing a small frequency shift. This allows us to perform the unitary transformation 
$T_1$=$\exp(-\mathrm{i}\nu ta^{\dagger}a/2)$
and switch to the quasi-interaction picture
$H_{I}=-Ka^{\dagger}a/2 + \mathrm{i}\chi(t)[a^{\dagger 2}\exp\left(\mathrm{i}t\nu\right)-a^{2}\exp\left(-\mathrm{i}t\nu\right)]$.
After applying the rotating wave approximation we obtain the
time-independent Hamiltonian~\cite{DodonovOneAtom}:
$H=({\rm i}\epsilon\omega_0/4)(a^{\dagger 2}-a^2)-Ka^\dagger a/2$.
Then, for the sake of simplicity, we move to a $\pi/4$ rotated reference frame
generated by means of the transformation $T_{2}=\exp(-{\rm i}\pi a^{\dagger}a /4)$,
which yields
\begin{align}\label{hamil_final}
\mathcal{H}=-{(\epsilon\omega_0/4)}\big(a^{\dagger 2}+a^2\big)-(K/2)a^\dagger a.
\end{align}

Equation~\eqref{hamil_final} represents the simplest effective Hamiltonian for which the photon production in the DCE can exhibit a threshold due to the off-resonance condition provided by the $K$ frequency shift. Now, by inserting $\mathcal{H}$ into the Schr\"odinger equation, $\mathrm{i}\frac{\mathrm{d}}{\mathrm{d}t}\ket{\Psi(t)}=\mathcal{H}\ket{\Psi(t)}$, and expanding the state vector $|\Psi(t)\rangle$ in terms of Fock states, $|\Psi(t)\rangle$=$\sum_{m=0}^\infty\mathcal{A}_m(t)|m\rangle$, we readily obtain an infinite set of coupled differential equations for
the transition probability amplitudes $\mathcal{A}_m(t)$:
\begin{align}\label{ampl_prob_eqs}
{\rm i}\mathcal{\dot{A}}_m(t)+\lambda_{m}\mathcal{A}_{m-2}(t)+\lambda_{m+2}\mathcal{A}_{m+2}(t)+
{K} m\mathcal{A}_m(t)/2=0,
\end{align}
where $\lambda_{m}$=$(\epsilon\omega_0/4)\sqrt{m(m-1)}$.
The solution of Eq.~\eqref{ampl_prob_eqs} is given by the
matrix elements  $\langle m|U(t)|\Psi(0)\rangle=\mathcal{A}_m(t)$,
where $U(t)$=$\exp\left(-{\rm i} \mathcal{H}t\right)$ and $|\Psi(0)\rangle$ is an initial pure state.
To compute $U(t)$ it is convenient to disentangle the exponential operator $\exp\left(-\mathrm{i} \mathcal{H}t\right)$. To do so, we introduce the operators $L_{+}$=$a^{\dagger 2}/2$, $L_{-}$=$a^2/2$ and $L_0$=$a^\dagger a/2+1/4$. Then, by computing the commutators, $[L_-,L_+]=2L_0$ and $[L_0,L_{\pm}]=\pm L_{\pm}$, we see that they close an algebra. Consequently, we can split the evolution operator, up to a global phase $\rm{e}^{-\rm{i}Kt/4}$, as~\cite{Ban93}
$U(t)$=$\beta_0^{{1}/{4}}\exp\big(\beta a^{\dagger 2}\big)\exp\big(a^\dagger a\ln\beta_{0}\big)
\exp\big(\beta a^2\big)$,
with $\beta$=${\rm i}\beta_0^{{1}/{2}}(1/2 \eta)\sinh\left(\eta\epsilon\omega_0t/2\right)$,
$\eta$=$\sqrt{1-(K/\epsilon\omega_0)^2}$
and
$\beta_0$=$[\cosh\left(\eta\epsilon\omega_0 t/2)-{\rm i}(K/\epsilon\omega_0\eta\right)\sinh\left(\eta\epsilon\omega_0 t/2\right)]^{-2}$.
Once we have written the evolution operator as a pro\-duct of exponentials,
we can readily evaluate its action over any initial state.

To estimate the photon production (average photon number) from the
vacuum, we compute the expectation value $\langle a^\dagger a\rangle_0=
\sum_{m=0}^\infty m|\mathcal{A}_m(t)|^2=
\langle 0|U^\dagger(t)a^\dagger aU(t)|0\rangle$=$
-{4\beta^2}{\beta_0^{-1}}$,
which yields the closed-form expression~\cite{DodonovEcGen}
\vspace{-0.2cm}
\begin{align}\label{vacuum_photons}
\langle a^\dagger a\rangle_0=\sinh^2\left(\eta \epsilon\omega_0t/2\right)/\eta^{2}.
\end{align}
Interestingly, Eq.~\eqref{vacuum_photons} exhibits three regimes de\-pen\-ding on
whether the ratio $K/\epsilon\omega_0$ is less, greater or equal to one. In the case where  $K/\epsilon\omega_0<1$, the photon ge\-ne\-ra\-tion grows exponentially. This is a pure manifestation of the quantum vacuum fluctuation amplification, which in the particular case of $K=0$ yields the well known expression $\langle a^\dagger a\rangle_0|_{K=0}=\sinh^2\left(\epsilon\omega_0t/2\right)$ ~\cite{Klimov1,Soff2004}.
In contrast, for $K/\epsilon\omega_0>1$, the photon production becomes oscillatory, va\-ni\-shing at
$\epsilon\omega_0t/2=n\pi/\sqrt{(K/\epsilon\omega_0)^2-1}$, with $n\in\mathbb{N}$.
Thirdly, at $K/\epsilon\omega_0=1$ the photon production is quadratic 
$\langle a^\dagger a\rangle_0=\left({\epsilon\omega_0}t/2\right)^2$, indicating a threshold between the exponential and the oscillatory behavior. It is noteworthy that at the threshold the photon production turns out to be half of the total number of photons  $\left(\mathcal{N}_{tot}\right)$ expected for the same system when the single mode approximation is not applied~\cite{DodonovReview}.
This result implies that, at threshold, 
our system governed by Eq.~\eqref{hamil_final} would produce a total photon growth as if it were governed by the exact Hamiltonian [Eq.~\eqref{:ham:eff:fock}] at the resonance condition.
%
\begin{figure}[t!]
\includegraphics[width=\linewidth,height=5cm]{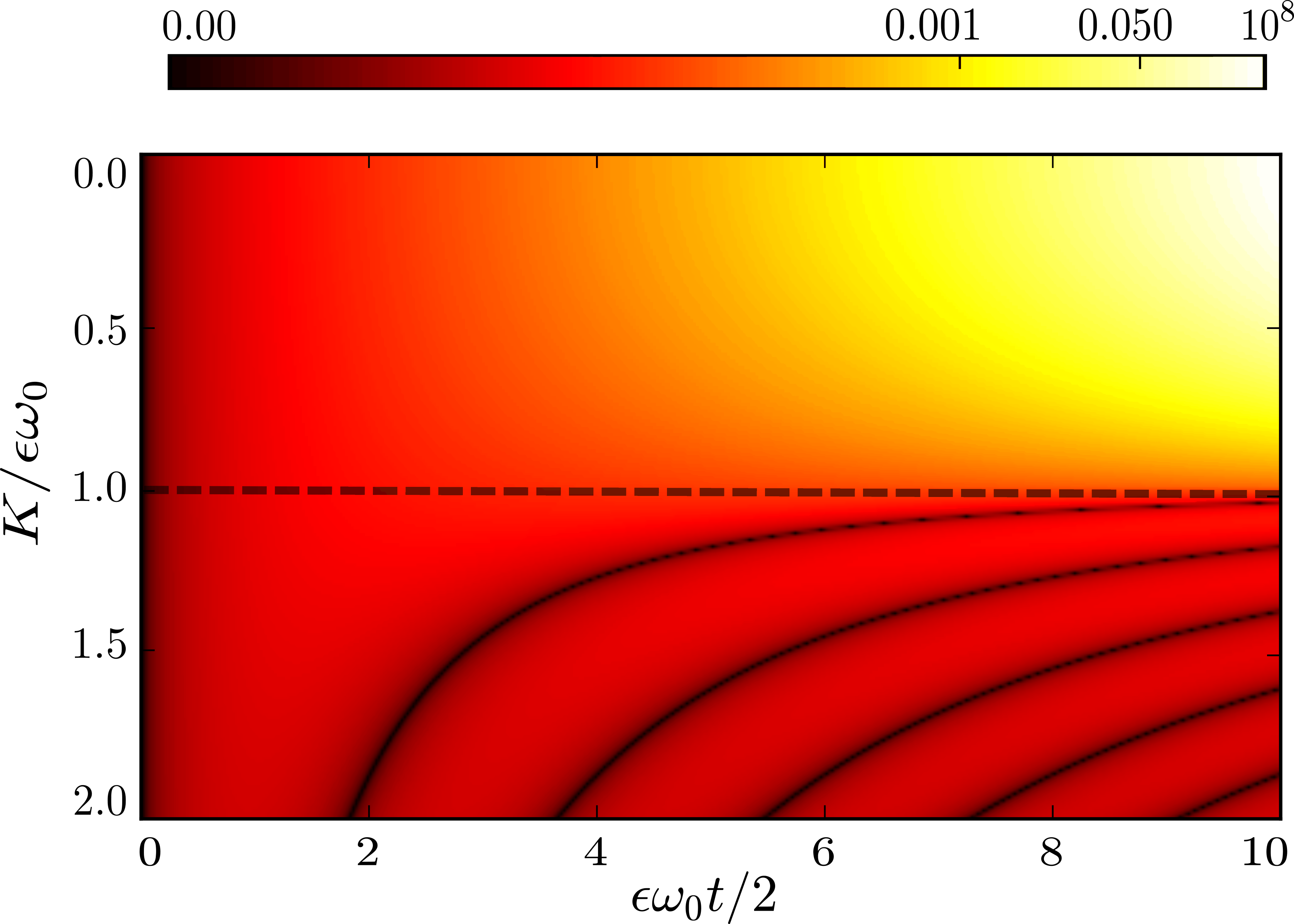}
\caption{
Photon generation from the vacuum state, $\langle a^\dagger a\rangle_0$,
produced by the vibrating cavity of Fig.~\ref{vibrating_cavity} (a).
The dashed line indicates the threshold of the system which
separates the exponential photon growth (above) from the oscillatory behavior (below).
Solid lines show the values where the photon production vanish. Colors are in log scale.}
\label{phase_diagram}
\vspace{-0.6cm}
\end{figure}

Figure~\ref{phase_diagram} shows a landscape of the photon production $\langle a^\dagger a\rangle_0$ evaluated in the three different regions. The threshold is marked by a dashed line that separates the exponential from the oscillatory behavior.
To understand these e\-ffects we explore the spectrum of the system. In particular, for $K/\epsilon\omega_0>1$, we see that $\mathcal{H}$ can be diagona\-lized using the transformation $S(r)$=$\exp[r(a^{\dagger 2}$-$a^2)/2]$, with $r$=$\frac{1}{4}\ln\left[\left(K-\epsilon\omega_0\right)/\left(K+\epsilon\omega_0\right)\right]$. Hence, the transformation $2S^\dagger(r)\mathcal{H}S(r)$= $-\epsilon\omega_0\sqrt{\left(K/\epsilon\omega_0\right)^{2}-1}\big(a^\dagger a$+$1/2\big)+K/2$ clearly indicates an equally spaced spectrum. This implies that any initial state is expected to undergo periodic revivals at particular instants given by the zeros of the function describing the photon production (average photon number). As depicted in Fig.~\ref{phase_diagram} the zeros are elucidated by the gaps (dark areas). On the other hand, for $K/\epsilon\omega_0<1$ the system exhibits a continuous spectrum which results in a significant photon production. Finally, at the threshold, all the eigenvalues coalesce.

{\em Simulation of DCE in photonic lattices.}
To translate concepts of the DCE 
to the optical domain we map the matrix elements of the $\mathcal{H}$-operator
over the inter-channel couplings and propagation constants of engineered waveguide
arrays, see Fig.~\ref{vibrating_cavity}(b). Within the nearest-neighbor regime the
normalized mode field amplitudes $\{\mathcal{E}_n(z)\}_{n=0}^\infty$
are governed by the set of
equations ~\cite{Sukhorukov1,Langari1,Perez_Glauber,Perez_Displaced_Fock}:
\begin{equation}\label{ampl_field_eqs}
{\rm i}{{\rm d}\mathcal{E}_n(z)}/{{\rm d}z}+\mathcal{C}_{n}\mathcal{E}_{n-1}(z)+
\mathcal{C}_{n+1}\mathcal{E}_{n+1}(z)+\alpha n\mathcal{E}_n(z)=0,
\end{equation}
where $z$ represents the propagation distance. Notice that Eq. \eqref{ampl_field_eqs} represents a tight-binding model with non-uniform site energies given by $\alpha n$, and hopping rates $\mathcal{C}_{n}$. To establish a one-to-one connection between the field amplitudes in the waveguide system, Eq.~\eqref{ampl_field_eqs}, and the probability amplitudes described by Eq.~\eqref{ampl_prob_eqs}, we define the coupling coefficients to be $\mathcal{C}_n=\mathcal{C}_1\sqrt{2n(2n-1)}$, for $n\geq 0$, where $\mathcal{C}_1$ stands for the coupling between the $0$th and $1$th waveguide. Moreover, the site energies $\alpha n$ correspond to the waveguide propagation constants obeying a transverse ramped refractive index~\cite{Perez_Bloch, Pertsch,Morandotti,Longhi1,Lebugle,Wang}. Notice that in a real waveguide array, the evanescent coupling between sites $n$ and $n-1$, separated by a distance $d_n$, is given by $C_{n}=C_1\exp[-(d_n-d_1)/s]$, with $d_1$ and $s$ being parameters of $\mathcal{C}_1$ that depend on the waveguide width and the associated optical wavelength ~\cite{Perez_Displaced_Fock,SzameitTutorial}.

For the system considered here, the full state-space representation is the harmonic oscillator space divided into even and odd subspaces. However, due to the quadratic nature of $\mathcal{H}$, the equations of motion for the system only connect states with the same parity. Since we are interested in the evolution of the system prepared in the vacuum state (an even state), the wave\-guide array shown in Fig.~\ref{vibrating_cavity}(b) is the proper one to si\-mulate the dynamics of the DCE, provided that $\mathcal{C}_1\rightarrow \epsilon\omega_0/4$ and $\alpha\rightarrow {K/2}$, with $z$ playing the role of time. Under these premises Eq.~\eqref{ampl_field_eqs} and the even terms of Eq.~\eqref{ampl_prob_eqs} are equivalent. Consequently, we can map second-neighbor interactions, such as in the DCE, on the nearest-neighbor interactions of a photonic lattice~\cite{Sukhorukov1}. Conversely, to simulate the odd terms of Eq.\eqref{ampl_prob_eqs}, one would require an independent photonic lattice, with the difference that only the odd terms should be considered. Indeed, the full system could be simulated by designing both arrays, one on top of the other, with a sufficient separation to neglect possible interactions between them~\cite{Langari1}.

To perform the DCE's photonic simulation, we excite the first waveguide (corresponding to $\ket{0}$ and labeled as $0$ in the waveguide array) with classical light. Accordingly, the field amplitude at site $m$, $\mathcal{E}_m(z)$, can be obtained by using the evolution operator discussed in the previous section $\mathcal{E}_m(z)=\langle 2m|U(z)|0\rangle$. By doing so, we find the intensity at the $m$-th waveguide
\vspace{-0.2cm}
\begin{align}\label{output_intensity}
I_{m}(z)= |\mathcal{E}_{m}(z)|^2 = \frac{(2m)!}{(2^{m}m!)^2}
\frac{\langle a^\dagger a\rangle^m_0}{\big(1+{\langle a^\dagger a\rangle_0}\big)^{m+\frac{1}{2}}},
\end{align}
which resembles the probability distribution for a thermal state. In this optical context, \emph{the photon production} $\langle a^\dagger a\rangle_0$ is expressed in terms of the ramping parameter $\alpha$ and the first coupling coefficient $C_{1}$ as $\langle a^\dagger a\rangle_0=\sinh^2\big(2C_1z\eta_x\big)\eta_x^{-2}$, with $\eta_x^2=1-x^2$ and $x=\alpha/2C_1$. Alternatively, one can write the photon production in terms of the intensity at the \emph{m}-th wave\-gui\-de as $\langle a^\dagger a\rangle_{0}^{\rm clas}=2\sum_{m=0}^N mI_m(z)$. This last expression constitutes the classical analogue of photon generation from the vacuum state at a distance $z$, where the factor 2 comes from considering only the even states, and $N$ is the maximum number of waveguides.

\begin{figure}[t!]
\includegraphics[width=\linewidth,height=7cm]{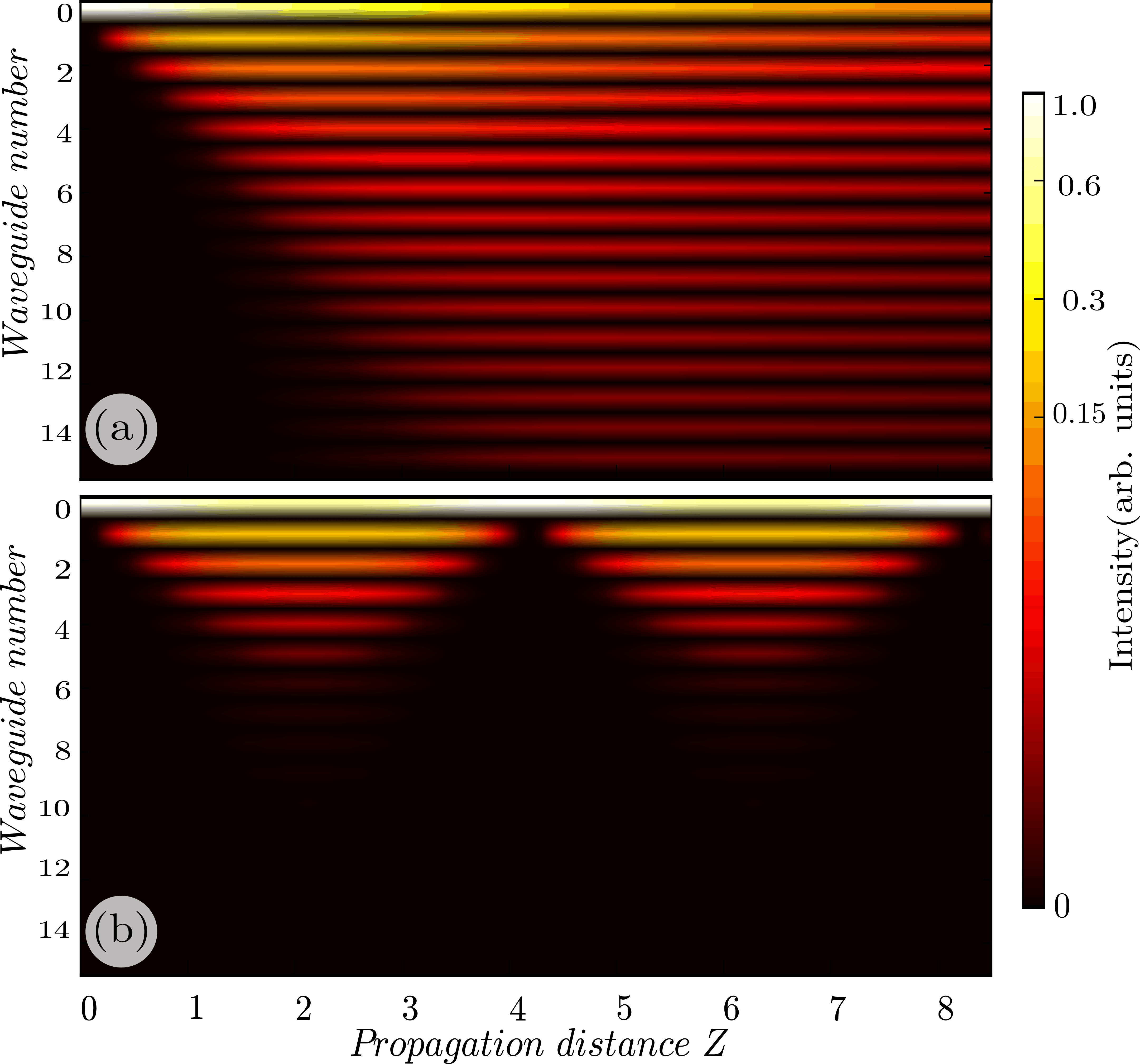}
\caption{
Light propagation in the waveguide array depicted in
Fig.~\ref{vibrating_cavity}(b) when the first site (vacuum state) is initially excited;
ramping constant $\alpha$=$0.5\mathrm{cm}^{-1}$.
(a) Extended propagation of the initial excitation, ballistic diffusion
representing a metal phase with $C_1$=$0.26\mathrm{cm}^{-1}$.
(b) Localization of the excitation (insulator phase)
with $C_1$=$0.2\mathrm{cm}^{-1}$, and first Bloch-like revival at $Z_{\rm rev}\sim 4$.}
\label{lattice_photon_vacum}
\vspace{-0.4cm}
\end{figure}

Figure~\ref{lattice_photon_vacum} shows the 
intensity distributions $I_m(z)$ for two waveguide arrays---designed using realistic experimental
parameters~\cite{Stutzer:17,Perez_Bloch}---above and below the threshold. For $x<1$ [Fig.~\ref{lattice_photon_vacum}(a)],
the propagation of the initial excitation, as a function
of the scaled distance $Z=2C_1z$, shows a rapid delocalization throughout the array.
In contrast, for $x>1$ [Fig.~\ref{lattice_photon_vacum}(b)] the system spectrum forms a Wannier-Stark ladder~\cite{Stutzer:17} causing spatial mode localization over some waveguides and giving rise to Bloch-like revivals at $Z_{\rm rev}$=$n\pi/\sqrt{x^2-1}$. Indeed, such spectral changes between an extended and localized excitation resemble a kind of metal-insulator phase transition~\cite{Langari1}, with a threshold at $x$=$1$. In that sense, Fig.~\ref{lattice_photon_vacum}(a) represents the metal phase, while Fig.~\ref{lattice_photon_vacum}(b) illustrates the insulator phase. Importantly, it should be stressed that photons produced in the metallic phase of the DCE can be easily detected, while their corresponding observation in the insulator phase is very challenging. Therefore, it is of interest to envision new approaches that may allow us to access the insulator phase of the DCE.

{\em Enhancing photon production through noise}.
As discussed above, when the DCE system is at the \emph{insulator phase}, the initial excitation remains localized around the first wave\-guide, i.e., the vacuum state. This results in a low photon production that could be measured only at very specific distances. To overcome this situation, we introduce a pure-dephasing mechanism, which simulates the effect of a Markovian environment interacting with the photonic system. This interaction results in the delocalization of the excitation, thus leading to an enhancement of photon production, a phenomenon called environment-assisted quantum transport \cite{Aspuru1,plenio2008,viciani2015, roberto2015-2, biggerstaff2015,roberto2017}. Indeed, to implement such mechanism in a waveguide array, one needs to include Gaussian fluctuations in the propagation constant of individual waveguides (dynamic diagonal disorder). This can be experimentally implemented by randomly changing the speed at which each waveguide is inscribed, see Refs. \cite{caruso2015,Robert_lattices} for details on the fabrication of such system. Remarkably, in the context of a non-stationary cavity field mode, this phenomenon could be observed by adding stochastic fluctuations to the frequency shift $K$.

Under different considerations over the random fluctuations, it entails to consider a pure-dephasing process for the field density operator $\rho$ ~\cite{Aspuru1,eisfeld2012,Robert_PRL}. Thus, we can study the action on $\rho$ of the generator $D[x]$ defined in the Lindblad form as $D[x]\rho=x\rho x^\dagger-(x^\dagger x\rho+\rho x^\dagger x)/2$, for which the master equation to solve takes the form
$\dot{\rho}=-{\rm i}[\mathcal{H},\rho]+\gamma D[a^\dagger a]\rho$, where $\gamma$ is the dephasing rate.
By inspecting the term $\langle n| D[a^\dagger a]\rho|m\rangle=-(n-m)^2\rho_{n,m}/2$, we can see that only the non-diagonal elements of the density matrix are directly affected by the random fluctuations, a footprint of the pure-dephasing process.
\begin{figure}[t!]
\includegraphics[width=\linewidth,height=9cm]{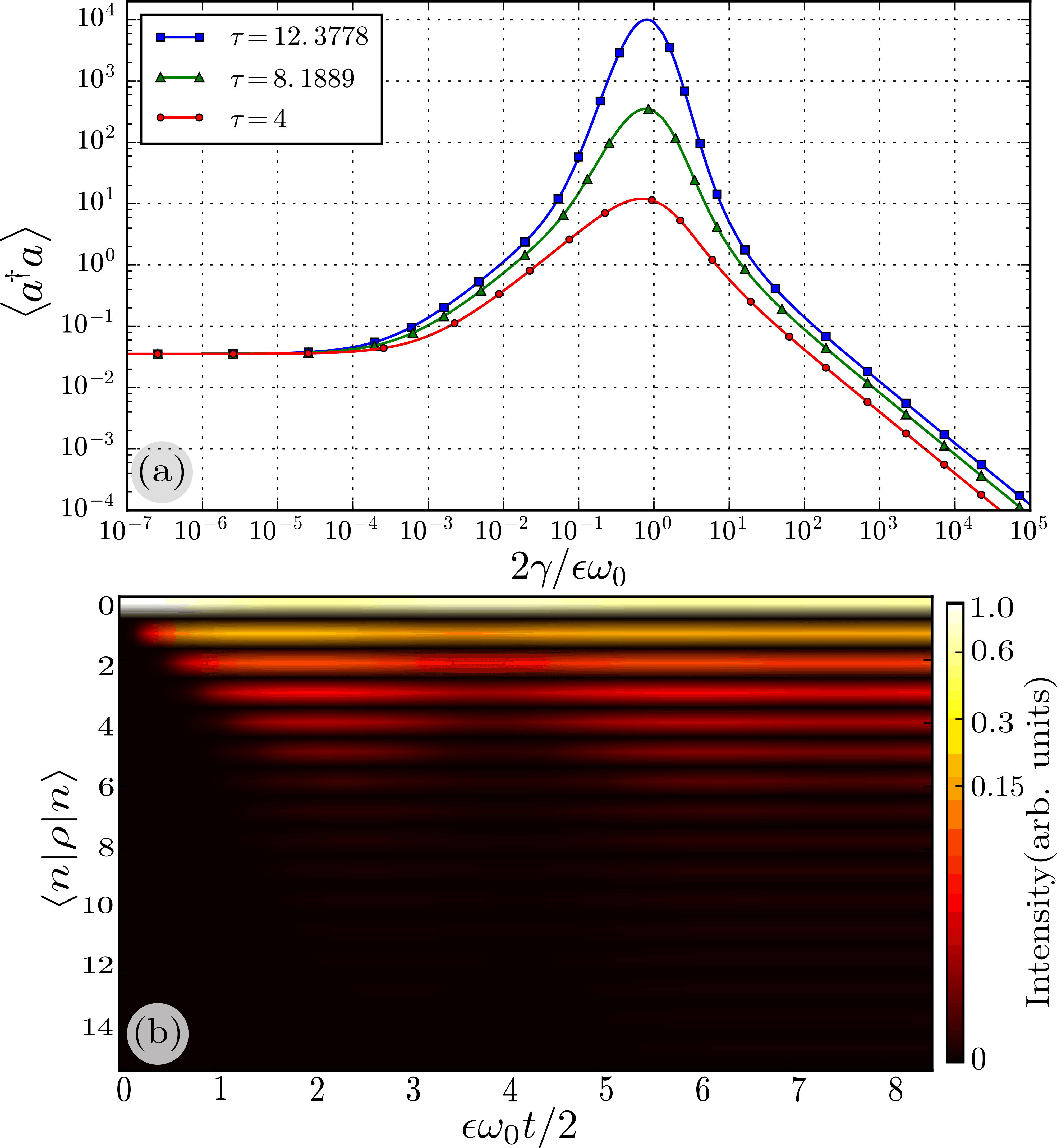}
\caption{(a) Enhancement of photon generation from the vacuum state as a function of the scaled dephasing rate, $2\gamma/\epsilon\omega_0$, at three revival times. (b) Evolution of the diagonal elements of the system's density matrix with $2\gamma/\epsilon\omega_0=0.04$.}
\label{increase_photons_production}
\vspace{-0.4cm}
\end{figure}


We now compute the average value of the number
and quadratic field operators. From the master equation, one can obtain the corresponding equations of motion: 
${\rm d}\langle {a^\dagger a} \rangle/{{\rm d}\tau}$
=${\rm i}(\langle a^{\dagger 2}\rangle-\langle a^{2}\rangle)$,
${\rm d}\langle {a^2}\rangle/{{\rm d}\tau} =2\tilde{K}_\gamma\langle {a}^2\rangle+
{\rm i}(2\langle a^\dagger a \rangle+1)$ and
${\rm d}\langle a^{\dagger 2}\rangle/{{\rm d}\tau}$=${\rm d}\langle {a^2}\rangle^*/{{\rm d}\tau}$,
%
where $\tau$=$\epsilon\omega_{0} t/2$ and $\tilde{K}_\gamma$=
${\rm i}K/\epsilon\omega_0-2\gamma/\epsilon\omega_0$.
Notice that the average photon number depends indirectly
on $\gamma$ through the quadratic field operator
expectation values. By considering the initial conditions
$\langle a^\dagger a\rangle|_{\tau=0}$=$
\langle a^2\rangle|_{\tau=0}$=$\langle a^{\dagger 2}\rangle|_{\tau=0}$=$0$
for the vacuum state, we have numerically solved the
above equations of motion. Figure~\ref{increase_photons_production}(a) shows the photon generation as a function of the dephasing rate at the insulator phase near the Bloch-like oscillations, that is, those regions where the photon production is minimum. We observe a significant improvement in the photon production for moderate values of the dephasing rate $\gamma$, a hallmark of noise-assisted transport ~\cite{Aspuru1,Robert_PRL}. This enhancement can be explained by looking at the dia\-go\-nal e\-le\-ments of the density matrix $\langle n|\rho|n\rangle$, as a function of time, for the initial  vacuum state, shown in Fig.~\ref{increase_photons_production} (b). Owing to the interaction of the system with a large environment, the delocalization of the excitation from the first site is broken, thus creating a larger contribution to the photon generation. Notice that delocalization is more dramatic at the Bloch-like revival regions, where photon production is zero if the system is not affected by noise. We have solved the master equation numerically using the QuTip package~\cite{Qutip}

%
%
%
{\em DCE at finite temperature}.-
Another mechanism for the enhancement of photon production is by considering thermal effects on the creation of photons. Indeed, it has been shown that photon generation in the DCE can be enhanced by several orders of magnitude depending on the wavelength and the temperature of the thermal radiation field that is used as initial state~\cite{Finite_Temperature}. To include thermal effects in our system, we make use of the Hamiltonian in Eq.~\eqref{hamil_final} to write the average photon number evolution for
an initial thermal field $\rho_{\rm th}(0)=\sum_{n=0}^\infty P_n |n\rangle\langle n|$, where
$P_n=\bar{n}_{\rm th}^n/(1+\bar{n}_{\rm th})^{n+1}$, and
$\bar{n}_{\rm th}=1/\big(\exp\left(\hbar\omega/k_BT\right)-1\big)$,
\vspace{-0.04cm}
\begin{align}\label{termal_photons}
\langle a^\dagger a\rangle_{\rm th}=\small{\sum}_{n=0}^\infty P_n\langle a^\dagger a\rangle_n=
(1+2\bar{n}_{\rm th})\langle a^\dagger a \rangle_0+\bar{n}_{\rm th}.
\end{align}
Here $\langle a^\dagger a\rangle_n =\langle n|U^\dagger(z) a^\dagger a U(z)|n\rangle = (1$+$2n)\langle a^\dagger a\rangle_0+n$ stands for the photon production from an initial Fock state $|n\rangle$. Notice that Eq.~\eqref{termal_photons} predicts an enhancement of photon production that depends on the average photon number of the thermal field. In the photonic context, a thermal field can be designed by making use of an independent disordered waveguide array, where an initially injected coherent state is thermalized, resulting in an incoherent mixture~\cite{Saleh2015}. Because the implementation of both systems relies on the use of the same integrated-optics technology, they could be designed in tandem, so the losses in the coupling of the thermal state to our proposed system would be negligible.
%
%

{\em Conclusions.-}
In this work we have proposed a novel photonic system for the classical simulation of
the dynamical Casimir effect. By using this system, we demonstrate that photon generation from the vacuum state may exhibit a transition from
an exponential growth to an oscillatory behavior, thus resembling a metal-insulator-like phase transition. Furthermore, we showed that the insulator phase appears as a result of the Bloch-like oscillations in the dynamics of the system. This causes a strong localization of the initial excitation in the first waveguide, leading to a poor contribution to the
Casimir-like radiation. To overcome this situation, we discussed two possible solutions. Firstly, we made use of a dephasing mechanism in which the coherent evolution of the system is broken by means of its interaction with a Markovian environment. The reduced coherence of the system causes a delocalization of the excitation, thus increasing the photon production by up to two orders of magnitude in the first Bloch-like revival. The second mechanism is based on the consideration of thermal effects in the DCE, we showed that in this case the photon production is enhanced by a factor that depends on the average photon number of the thermal field that is initially injected in the system. Finally, we would like to point out that DCE modifications due to Kerr nonlinearities, recently predicted in Ref.~\cite{ricardo2017}, could also be tested in our proposed photonic system by changing the linear refractive index profile to a quadratic one.

{\em Acknowledgments}.-
R.R.-A. thanks CONACYT, Mexico, for Scholarship No. 379732 and
Project No. 166961, and DGAPA-UNAM, Mexico,
for support under Project No. IN108413.

%
%
%
\bibliography{casimir_dinamico}
\end{document}